# False negatives for remote life detection on ocean-bearing planets: Lessons from the early Earth

Christopher T. Reinhard[1,2], Stephanie L. Olson[1,3], Edward W. Schwieterman[1,3,4,5] Timothy W. Lyons[1,3]

[1]NASA Astrobiology Institute
[2]School of Earth & Atmospheric Sciences, Georgia Institute of Technology
[3]Department of Earth Sciences, University of California, Riverside
[4]NASA Postdoctoral Program, Universities Space Research Association, Columbia, Maryland, USA
[5]Blue Marble Space Institute of Science, Seattle, WA

**ABSTRACT**

Ocean-atmosphere chemistry on Earth has undergone dramatic evolutionary changes through its long history, with potentially significant ramifications for the emergence and long-term stability of atmospheric biosignatures. Though a great deal of work has centered on refining our understanding of *false positives* for remote life detection, much less attention has been paid to the possibility of *false negatives*, that is, cryptic biospheres that are widespread and active on a planet's surface but are ultimately undetectable or difficult to detect in the composition of a planet's atmosphere. Here, we summarize recent developments from geochemical proxy records and Earth system models that provide insight into the long-term evolution of the most readily detectable potential biosignature gases on Earth – oxygen ($O_2$), ozone ($O_3$), and methane ($CH_4$). We suggest that the canonical $O_2$-$CH_4$ disequilibrium biosignature would perhaps have been challenging to detect remotely during Earth's ~4.5 billion year history and that in general atmospheric $O_2$/$O_3$ levels have been a poor proxy for the presence of Earth's biosphere for all but the last ~500 million years. We further suggest that detecting atmospheric $CH_4$ would have been problematic for most of the last ~2.5 billion years of Earth's history. More broadly, we stress that internal oceanic recycling of biosignature gases will often render surface biospheres on ocean-bearing silicate worlds cryptic, with the implication that the planets most conducive to the development and maintenance of a pervasive biosphere will often be challenging to characterize via conventional atmospheric biosignatures.

## 1. Introduction

In the two decades since the first radial velocity surveys detected distant planets orbiting Sun-like stars (Mayor and Queloz, 1995), the burgeoning field of exoplanet research has yielded an astonishing number and diversity of extrasolar worlds. At the time of this writing, there are 2,331 confirmed exoplanets, with an additional 2,365 candidate planets in the Kepler field of view (Akeson *et al.*, 2013), many of which will likely be confirmed as additional exoplanets. The size, orbital parameters, and composition of these planets are staggering in their diversity (e.g., Pierrehumbert, 2013), and the continued discovery and characterization of these worlds may ultimately have the potential to yield our first evidence for life beyond Earth.

Despite the anticipation that detection and analysis of exoplanet atmospheres would be difficult or impossible to perform routinely, over three dozen exoplanet atmospheres have been observed to date (Seager and Deming,

2011; Seager, 2013). The next generation of space telescopes, including the James Webb Space Telescope (JWST) and Transiting Exoplanet Survey Satellite (TESS), are expected to detect and/or characterize a number of transiting exoplanets in unprecedented detail (see, for example, Gardner *et al.*, 2006; Deming *et al.*, 2009; Cowan *et al.*, 2015). Large space-based telescope missions currently in their science definition phase would possess the capability to directly image terrestrial exoplanets at UV to near-infrared wavelengths (e.g., Dalcanton *et al.*, 2015; Mennesson *et al.*, 2016), while future ground-based observatories will also be able to spectrally characterize the atmospheres of small planets around the very nearest stars (e.g., Snellen *et al.* 2013; 2015). Indeed, for the foreseeable future our only accessible method for detecting life or even fully characterizing habitability beyond Earth will likely be deciphering the chemistry of exoplanet atmospheres (Meadows and Seager, 2010; Seager, 2014).

The prospect of remotely analyzing the composition of potentially habitable exoplanetary surfaces and atmospheres provides strong impetus for the development of biosignatures that can be used to diagnose the presence and scope of a surface biosphere. The most prominent of these approaches emphasize the potential of a planet's biosphere to reshape atmospheric chemistry and in particular to drive chemical disequilibrium via large production fluxes of incompatible volatile species (Lovelock, 1965; Hitchcock and Lovelock, 1967; Sagan *et al.*, 1993; Kaltenegger *et al.*, 2007; Meadows and Seager, 2010; Seager *et al.*, 2013a; Krissanson-Totton *et al.*, 2016). The most frequently cited example of this concept is the co-existence of molecular oxygen ($O_2$) and methane ($CH_4$) in the modern Earth's atmosphere at abundances that are many orders of magnitude out of thermodynamic equilibrium (e.g., Hitchcock and Lovelock, 1967), but this concept can be functionally extended to include, for example, the coexistence of $N_2$, $O_2$, and $H_2O$ in Earth's modern ocean-atmosphere system at levels strongly out of thermodynamic equilibrium with aqueous $H^+$ and $NO_3^-$ (Krissanson-Totton *et al.*, 2016).

Significant effort has been expended to understand abiotic mechanisms for generating the most prominent among the possible atmospheric biosignatures, particularly false positive signatures based on oxygen ($O_2$) and ozone ($O_3$). A number of 'pathological' high-$O_2$ scenarios are now known, including those that involve: (1) hydrogen escape from atmospheres depleted in non-condensing gases (Wordsworth and Pierrehumbert, 2014), (2) $CO_2$ photolysis in very dry atmospheres (Gao *et al.*, 2015), (3) atmospheres undergoing runaway water loss (Luger and Barnes, 2015), and (4) modest photochemical production of $O_2/O_3$ from $CO_2$ photolysis on planets around certain types of host star (Segura *et al.*, 2007; Domagal-Goldman *et al.*, 2014; Harman *et al.*, 2015). Although the diversity of mechanisms for generating high $O_2/O_3$ levels on lifeless worlds sounds an important cautionary note in the search for compelling biosignatures, these studies also identified a suite of contextual tools that can be used to diagnose high-$O_2$ false positives under many circumstances (Domagal-Goldman *et al.*, 2014; Harman *et al.*, 2015; Schwieterman *et al.*, 2015, 2016).

Biosignatures in reducing planetary atmospheres have been less explored (Domagal-Goldman *et al.*, 2011; Seager *et al.*, 2013b). This gap is important, as Earth's atmosphere has been relatively reducing – but not $H_2$-dominated – for most of its history (e.g., Lyons *et al.*, 2014), and it is currently unknown whether the evolution of water-splitting (oxygenic) photosynthesis should be expected to be a common phenomenon beyond our solar system. Indeed, even if energetic benefits make the evolution of oxygenic photosynthesis probable on habitable planets, this alone may not be enough. For example, Earth's atmosphere may have remained strongly reducing for 500 million years or more after the emergence of biological $O_2$ production (Catling and Claire, 2005; Planavsky *et al.*, 2014a). As a result, there is little *a priori* reason to expect that most life-bearing extrasolar worlds will be characterized by atmospheres rich in biogenic $O_2$, particularly given our intrinsically brief timescale of observation relative to the potential timescales of planetary oxygenation.

An additional issue that has not been fully explored is that atmospheric chemistry on ocean-bearing planets can be strongly decoupled from overall surface metabolic fluxes due to internal microbial recycling within a planet's oceanic biosphere. For example, some of the largest input/output fluxes in Earth's modern biological $CH_4$ cycle occur via the microbial production and oxidation of $CH_4$ in deep marine sediments (Reeburgh, 2007), and this cycle is entirely decoupled from Earth's atmospheric $CH_4$ inventory. Indeed, most of this cycling occurs through anaerobic processes in the ocean (Valentine, 2011) and can occur through a number of common electron acceptors that will not necessarily be in equilibrium with corresponding atmospheric gases (Martens and Berner, 1977; Hoehler *et al.*, 1994; Regnier *et al.*, 2011; Sivan *et al.*, 2011; Deutzmann *et al.*, 2014; Riedinger *et al.*, 2014). As a result, internal recycling by ocean biospheres may inhibit the development or erode the stability of atmospheric biosignatures despite large production rates of biosignature gases. Such oceanic filtering of atmospheric biosignatures could in principle occur on a wide range of ocean-bearing silicate planets, and in many cases will be an important process even on reducing planetary surfaces.

A second example is provided by the possibility of a cryptic oxygenic biosphere. In Earth's oceans, slow air-sea exchange of $O_2$ relative to high local rates of biological production in the surface waters can result in large regions of the surface ocean that are strongly out of equilibrium with atmospheric $pO_2$, by as much as a factor of $\sim 10^4$ (Kasting, 1991; Olson *et al.*, 2013; Reinhard *et al.*, 2013a; Reinhard *et al.*, 2016). Thus, despite vigorous biological $O_2$ production in the shallowest oceanic environments, a planetary atmosphere can remain pervasively reducing as a result of either low globally integrated surface fluxes or elevated atmospheric sinks (or some combination). Such planets may be capable of accumulating significant quantities of reduced biosignature gases but will be false negatives for the presence of oxygenic photosynthesis and will have limited capacity for preservation and detection of associated oxidized biosignatures gases (e.g., $O_3$, $N_2O$).

As a proof on concept, we briefly summarize the remote detectability of $O_2/O_3$, $CH_4$, and the $O_2$-$CH_4$ disequilibrium throughout Earth's history. Our analysis suggests that the $O_2$-$CH_4$ disequilibrium approach would have failed for most of Earth's history, particularly for observations at low to moderately high spectral resolving power ($R \leq 10,000$). In addition, it is possible that $O_2/O_3$ may only have been applicable as a potential biosignature during the last $\sim 10\%$ of Earth's lifetime. As a result, most of our planet's history may have been characterized by either high abundances of a single biogenic gas that can also have significant abiotic sources (e.g., $CH_4$) or by a cryptic biosphere that was widespread and active at the surface but remained ultimately unrepresented in the detectable composition of Earth's atmosphere. Finally, we argue that cryptic biospheres may be a particularly acute problem on ocean-bearing planets, with the implication that many of the most favorable planetary hosts for surface biospheres will also have high potential for attenuation of atmospheric biosignatures.

## 2. Evolution of atmospheric $O_2/O_3$ during Earth's history

The most compelling quantitative insights into atmospheric $O_2$ abundance during Archean time (~3.8-2.5 Ga) come from the observation of large non-mass-dependent (NMD) sulfur isotope anomalies in sulfide and sulfate minerals deposited in coeval marine sediments (Farquhar *et al.*, 2000; 2001; reviewed in Johnston, 2011). When interpreted in the context of photochemical models (Pavlov and Kasting, 2002; Ono *et al.*, 2003; Zahnle *et al.*, 2006; Ueno *et al.*, 2009), the presence of these isotopic anomalies implies a ground-level atmospheric $pO_2$ value that was below $10^{-5}$ times the present atmospheric level (PAL) and more likely below ~$10^{-7}$ PAL (e.g., Claire *et al.*, 2006; Zahnle *et al.*, 2006). Though there is some evidence for transient periods of elevated $pO_2$ during the Archean (Anbar *et al.*, 2007; Reinhard *et al.*, 2009; Kendall *et al.*, 2015), these deviations are poorly understood from a quantitative standpoint, and the overall background state during Archean time appears to have been one of exceptionally low atmospheric $O_2$. The sulfur isotope anomalies fingerprinting extremely low atmospheric $O_2$ disappear relatively rapidly during the "Great Oxidation Event" (GOE) of the early Paleoproterozoic (Holland, 2002; Luo *et al.*, 2016) and are not seen again in typical marine sedimentary rocks for the remainder of Earth's history (Fig. 1).

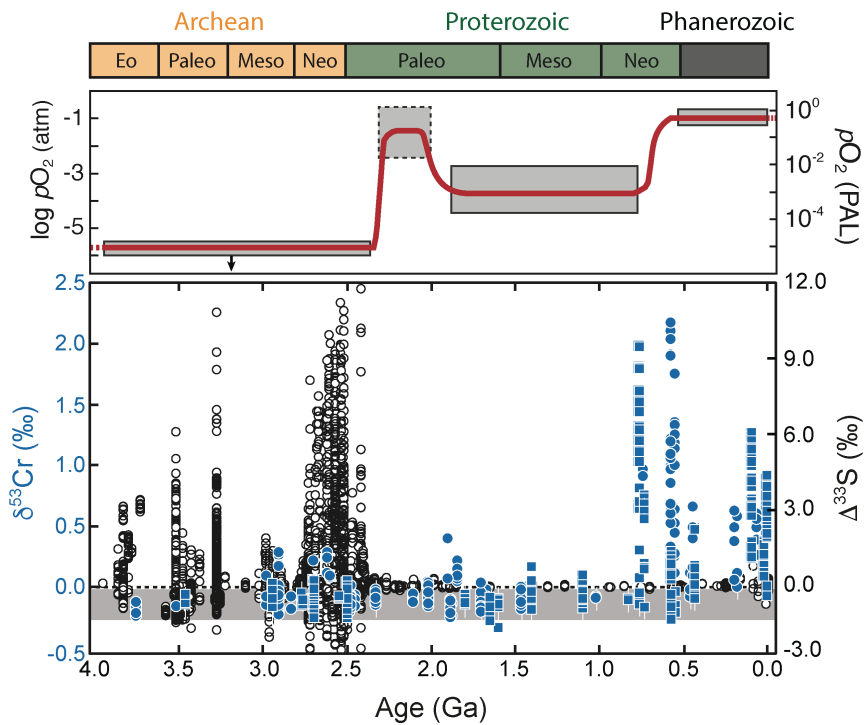

**Figure 1.** Atmospheric $O_2$ on Earth through time (upper) and select geochemical proxies for atmospheric $pO_2$ (lower). In the upper panel, shaded boxes show approximate ranges based on geochemical proxy reconstructions, while the red curve shows one possible trajectory through time. In the lower panel, open circles show the magnitude of non-mass-dependent sulfur isotope (NMD-S) anomalies, shown as $\Delta^{33}S$ (see Johnston, 2011) and compiled as in Reinhard et al. (2013b). Filled blue symbols show chromium (Cr) isotope fractionations from iron-rich and siliciclastic marine sedimentary rocks throughout Earth's history (Cole et al., 2016). We show these particular proxies in order to optically emphasize the three major periods of Earth's history discussed here: the Archean (~3.8-2.5 Ga), the Proterozoic (~2.5-0.5 Ga), and the Phanerozoic (~500 Ma to the present). See Lyons et al. (2014) for a more in-depth recent discussion of atmospheric $O_2$ proxies and evolution through time.

Unfortunately, the rather extreme sensitivity of the sulfur isotope system to very low levels of O₂ in the atmosphere makes it unfeasible to quantitatively diagnose changes in atmospheric composition once $pO_2$ increased above ~1-2 ppmv during the GOE. As a result, atmospheric O₂ levels for most of the subsequent ~2.5 billion years of Earth's history are rather poorly constrained. There is some evidence from the geochemical record for a transient, though protracted, rise in atmospheric $pO_2$ during the Paleoproterozoic (Schröder et al., 2008; Planavsky et al., 2012; Partin et al., 2013; Canfield et al., 2015), perhaps to levels that would have rivaled those of the modern Earth (Fig. 1). However, this evidence remains somewhat fragmentary at present and is difficult to evaluate quantitatively.

Nevertheless, a number of geochemical records point to low atmospheric $pO_2$ for extended periods of Earth's history well after the GOE. For example, the apparent loss of Fe (and in some cases Mn) as dissolved reduced species from paleosols (ancient weathering profiles) during some periods of the Proterozoic (Holland, 1984) and the general lack of significant chromium (Cr) isotope fractionation in marine sediments (Planavsky et al., 2014b; Cole et al., 2016) are both consistent with very low atmospheric $pO_2$ between ~1.8-0.8 Ga (Fig. 1). Such low $pO_2$ conditions are also echoed in the numerous indicators of low oxygen in the deep and shallow ocean during this period (Planavsky et al., 2011; Poulton and Canfield, 2011; Reinhard et al., 2013c; Tang et al., 2016). However, a smaller atmospheric O₂ inventory during this period carries with it the potential for somewhat unstable behavior (e.g., Planavsky et al., 2014), and there is proxy evidence for potential short-term fluctuations in atmospheric $pO_2$ between ~1.8-0.8 Ga (e.g., Sperling et al., 2014; Gilleaudeau et al., 2016).

Geochemical records during the Neoproterozoic (~1,000 – 541 Ma) suggest significant ocean-atmosphere redox shifts just prior to the Sturtian glacial epoch at ca. 720 Ma (Planavsky et al., 2014b; Thomson et al., 2015) followed by a potentially highly dynamic interval leading up to the Precambrian-Cambrian boundary at 541 Ma (Sahoo et al., 2016). Atmospheric $pO_2$ during the earliest Paleozoic (~543-400 Ma), though not particularly well constrained at present, was likely higher than that of the late Proterozoic but well below modern (Bergman et al., 2004; Dahl et al., 2010; Lyons et al., 2014). Thereafter, the charcoal record strongly suggests that $pO_2$ values during most of the last ~400 million years have been above ~50% PAL (e.g., Cope and Chaloner, 1980; Chaloner, 1989), making O₂ a dominant constituent of Earth's atmosphere for most of the last half-billion years (Fig. 1; Table 1).

From a practical perspective, the detection of O₂ can be achieved via proxy by searching for signs of O₃. On the modern Earth, stratospheric O₃ is ultimately derived from the photolysis of atmospheric O₂, which generates oxygen atoms that subsequently combine with O₂ molecules to yield O₃ (the first two of the so-called "Chapman reactions"):

$$O_2 + h\nu \rightarrow O + O \; (\lambda < 242nm) \quad ,$$
$$O + O_2 + M \rightarrow O_3 + M \quad ,$$

where $M$ is any inert molecule that can absorb the energy of the excited O₃ molecule following collision between O and O₂ and eventually dissipate it as heat. The column abundance of atmospheric ozone (O₃) shows a nonlinear dependence on the O₂ content of Earth's atmosphere in 1-D photochemical models (Kasting and Donahue, 1980). In addition, both the altitude and overall abundance of peak O₃ increase with ground-level atmospheric $pO_2$ (Fig. 2A).

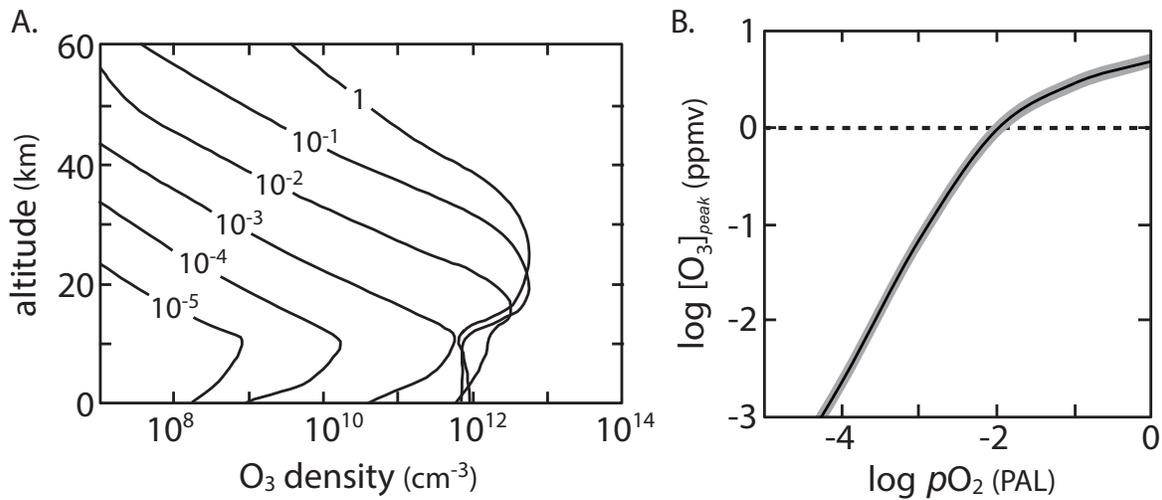

**Figure 2.** Shown in (A) are number densities for atmospheric ozone ($O_3$) as a function of altitude calculated from a 1-D photochemical model (after Kasting and Donahue, 1980) at different atmospheric $pO_2$ values (relative to the present atmospheric level, PAL). The calculations assume a solar zenith angle of 45°. Shown in (B) are calculated values of peak atmospheric $O_3$ as a function of ground-level $pO_2$ according to the results of Kasting and Donahue (1980). The calculations assume an atmospheric scale height of 7 km and a range of temperatures between 200-240K, but are not particularly sensitive to either of these values. The horizontal dashed line denotes a peak $O_3$ level of 1 ppmv (see text).

Considering this in light of the geochemical records discussed above, it is possible that for much of Earth's history atmospheric $O_3$ abundance was extremely low. We can estimate the effects of this by using the characteristics of $O_3$ profiles generated by a suite of 1-D photochemical models (Kasting and Donahue, 1980) to calculate peak $O_3$ levels in Earth's atmosphere as a function of ground-level $pO_2$ (Fig. 2B). The results of this exercise suggest that peak mid-Proterozoic atmospheric $O_3$ would have been ~0.1-1 ppmv based on the ground-level $pO_2$ values estimated for this time interval (e.g., $pO_2$ ~ 0.1-1% PAL) and many orders of magnitude below this during the Archean. However, during most of the Phanerozoic, and potentially some periods of the Paleoproterozoic, atmospheric $O_3$ would have been present at values approaching that of the modern Earth (Table 1).

## 3. Evolution of atmospheric $CH_4$ during Earth's history

Atmospheric $O_2/O_3$ abundances estimated based on the ground-level $pO_2$ values discussed above would have had an important impact on another key biosignature gas – methane ($CH_4$). The relationship between the atmospheric lifetime of $CH_4$ and the $O_2$ content of Earth's atmosphere is complex and somewhat counterintuitive. As atmospheric $pO_2$ drops from 1 PAL to $10^{-1}$ PAL, the atmospheric lifetime of $CH_4$ increases as a result of decreasing rates of $CH_4$ destruction in the atmosphere by hydroxyl radicals (OH). However, once $pO_2$ drops below ~$10^{-2}$ PAL the atmospheric lifetime of $CH_4$ decreases sharply. This latter behavior arises as a result of increased photochemical production of tropospheric OH as UV shielding by $O_3$ decreases (Kasting and Donahue, 1980; Pavlov et al., 2000; Pavlov et al., 2003) (Kasting & Donahue, 1980; Pavlov & Kasting, 2002; Pavlov et al., 2003). An additional complexity is that the atmospheric lifetime of $CH_4$ is also a highly non-linear function of its source fluxes from the biosphere and solid Earth (Pavlov *et al.*, 2003), which means that potential oceanic sinks for $CH_4$ must also be considered in any attempt to diagnose atmospheric $pCH_4$. Because the biospheric production and consumption fluxes of $CH_4$ are also directly or indirectly linked to atmospheric $O_2$ levels via the availability of oxidants at a planet's surface (e.g., $O_2$, $Fe^{3+}$, $SO_4^{2-}$), the quantitative relationship between atmospheric $O_2$ and $CH_4$ will be extremely complex on ocean-bearing silicate planets.

**Table 1.** Atmospheric $pO_2$ and $pCH_4$ values for different geologic epochs used in our spectral calculations.

| Age | | $pO_2$ (PAL) | | $pCH_4$ (ppmv) | |
| --- | --- | --- | --- | --- | --- |
| stratigraphic | Ma | low | high | low | high |
| Archean | 3800 – 2500 | -- | $10^{-5}$ | $10^2$ | $10^3$ |
| Lomagundi | 2200 – 2000 | $10^{-2}$ | 2.0 | 0.7 | 6 |
| mid-Proterozoic | 1800 – 800 | $10^{-3.3}$ | $10^{-2}$ | 1 | 20 |
| Phanerozoic | 543 – 0 | 0.25 | 1.4 | 0.7 | 6 |

Values for $pO_2$ are approximated according to Rye and Holland (1998), Pavlov and Kasting (2002), Bergman *et al.* (2004), Glasspool and Scott (2010), Crowe *et al.* (2013), Lyons *et al.* (2014), Planavsky *et al.* (2014), and Tang *et al.* (2016). Values for $pCH_4$ are approximated according to Pavlov *et al.* (2000; 2003), Claire *et al.* (2006), Zahnle *et al.* (2006), Catling *et al.* (2007), Bartdorff *et al.* (2008), and Olson *et al.* (2016).

Though there is at present no available geochemical proxy for ancient atmospheric $CH_4$ levels, a number of recent studies have attempted to explicitly model the effect of metabolic $CH_4$ consumption on surface fluxes and atmospheric $pCH_4$ (Claire *et al.*, 2006; Catling *et al.*, 2007; Beal *et al.*, 2011; Daines and Lenton, 2016; Olson *et al.*, 2016). The most recent attempt includes both aerobic and anaerobic microbial oxidation of $CH_4$ within a 3-D model of ocean biogeochemistry coupled to a parameterized $O_2$-$O_3$-$CH_4$ photochemical scheme (Olson *et al.*, 2016). Overall, the results of this work suggest that atmospheric $CH_4$ levels are extremely sensitive to concentrations of seawater sulfate ($SO_4^{2-}$). As a result, at atmospheric $pO_2$ values and marine $SO_4^{2-}$ levels characteristic of the Proterozoic, net biogenic $CH_4$ fluxes would have been strongly attenuated such that atmospheric $pCH_4$ was unlikely to have been significantly higher than ~1-10 ppmv (Fig. 3; Olson *et al.*, 2016). However, given recent estimates of Archean seawater $SO_4^{2-}$ levels (Crowe *et al.*, 2014), it is likely that biogenic $CH_4$ fluxes to the atmosphere would have been much higher during this period. Combined with much lower atmospheric $pO_2$, these larger fluxes could potentially have led to Archean atmospheric $pCH_4$ values on the order of ~$10^2$-$10^3$ ppmv (Pavlov *et al.*, 2000; Claire *et al.*, 2006; Zahnle *et al.*, 2006; Catling *et al.*, 2007).

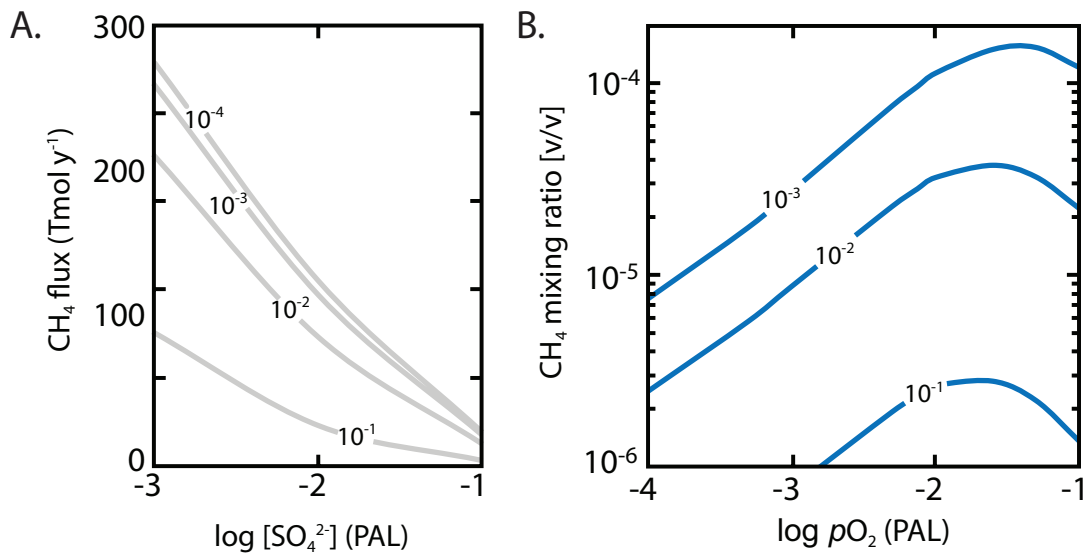

**Figure 3.** Representative results from an intermediate-complexity Earth system model of the ocean-atmosphere $O_2$-$O_3$-$CH_4$ cycle. Shown in (A) are net biospheric $CH_4$ fluxes to the atmosphere as a function of global average marine sulfate concentration ([$SO_4^{2-}$]), with contours labeled according to atmospheric $pO_2$ value (relative to the present atmospheric level, PAL). Shown in (B) are steady-state atmospheric $CH_4$ mixing ratios as a function of atmospheric $pO_2$, with contours labeled according to marine [$SO_4^{2-}$] (relative to the present oceanic level, POL). Simulations performed under the same default conditions as those in Olson et al. (2016).

## 4. Remote detectability of $O_2$-$O_3$-$CH_4$ biosignatures throughout Earth's history

The actual detectability of any gas in a future exoplanet observation will be a combined function of the abundance of that gas in the atmosphere, the observing parameters (such as distance, stellar host type, background noise, etc.), and the instrumentation with which the observation is made (see, e.g., Robinson et al., 2016). Here, we conduct a qualitative assessment of the potential detectability of $O_2$, $O_3$, and $CH_4$ through geologic time by generating synthetic direct-imaging spectra based on scaling a modern mid-latitude Earth atmosphere profile (Schwieterman et al., 2015) to the evolving gas abundances throughout Earth's history (Table 1). We use the Spectral Mapping Atmosphere Radiative Transfer Model (SMART; Meadows & Crisp, 1996; Crisp 1997), which is well validated by remote observations of the whole Earth (Robinson et al., 2011; 2014), Venus (Arney et al., 2014), and Mars (Tinetti et al., 2005), and has been previously used in spectral studies of the Archean Earth (Arney et al., 2016). We focus on the $O_3$ UV Hartley-Huggins bands (centered at ~0.25 μm), the $O_2$ Fraunhoffer A band (0.76 μm), and the 1.65 μm $CH_4$ band as a function of geologic epoch (Fig. 4). These bands were chosen in part because they will likely be included in the instrumentation suite of future direct-imaging telescopes (e.g., Dalcanton et al., 2015) and because MIR direct-imaging of exoplanets is unlikely for the intermediate future.

Molecular oxygen ($O_2$) has no significant spectral features at mid-infrared (IR) wavelengths, but it has three significant features in the optical range (Des Marais et al., 2002). These are the so-called "Fraunhofer" A and B bands (at 0.76 and 0.69 μm, respectively), along with an additional feature at 1.26 μm. Of these, the Fraunhofer A band is the most prominent but is likely to have appreciable depth only at atmospheric $pO_2$ levels of ~1% or higher (Fig. 4; Des Marais *et al.*, 2002). Thus, direct detection or quantification of molecular oxygen ($O_2$) would have been very difficult for much of Earth's history, other than during a transient period of the Paleoproterozoic and during most of the last ~500 million years (Fig. 1, 4).

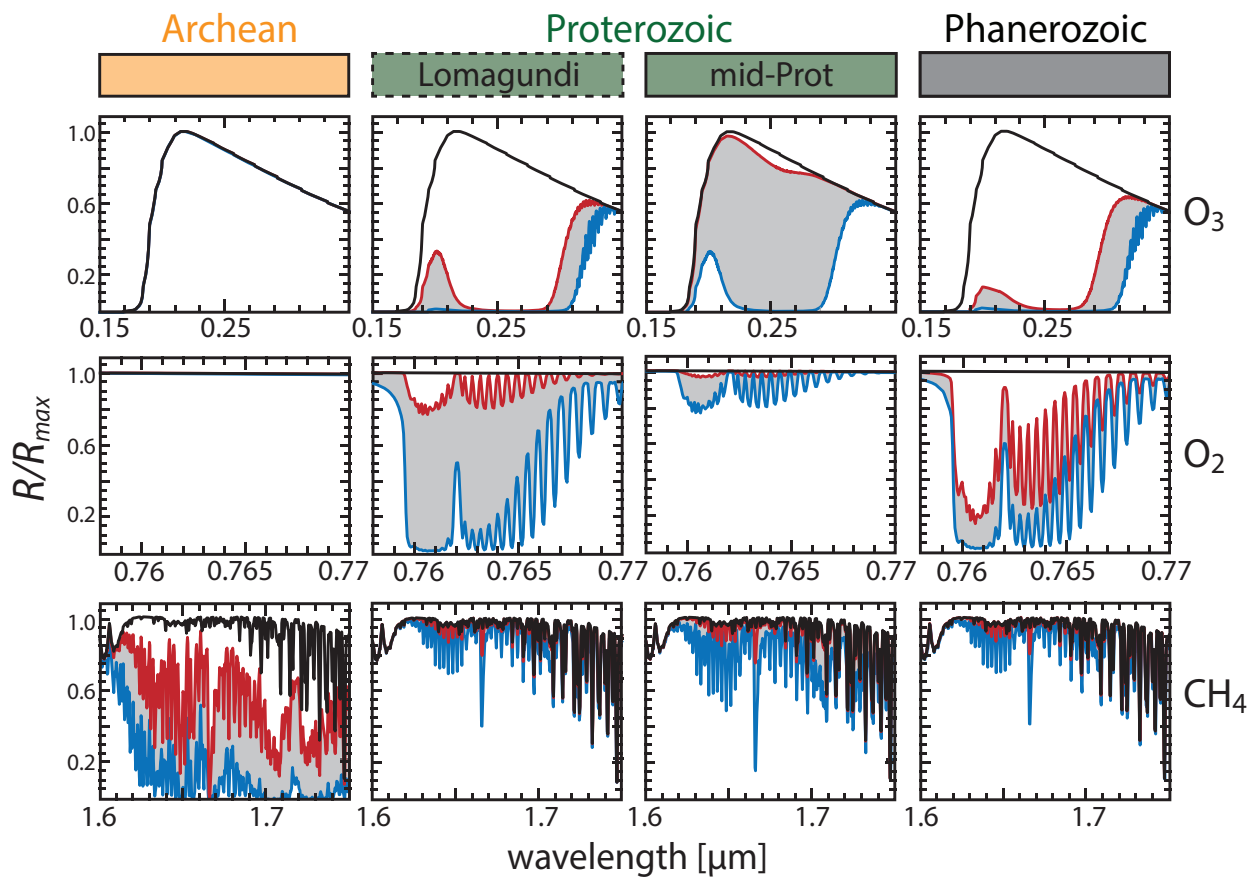

**Figure 4.** Reflectance spectra of selected $O_2$, $O_3$, and $CH_4$ bands as a function of geologic epoch. Lower abundance limits are given in red, upper limits are given in blue, and the region between these limits is shaded grey. The black line represents the case with no absorption of $O_2$, $O_3$, or $CH_4$. Limits are representative of both uncertainties in atmospheric abundance and the variability of those abundances over the course of each epoch (Table 1). Values for peak atmospheric $O_3$ are calculated as a function of ground-level $pO_2$ according to the results of Kasting and Donahue (1980). The resolution of each spectrum is 1 cm$^{-1}$, which is approximately $\Delta\lambda = 6.25 \times 10^{-6}$ µm at 0.25 µm, $\Delta\lambda = 5.78 \times 10^{-5}$ µm at 0.76 µm, and $\Delta\lambda = 2.72 \times 10^{-4}$ µm at 1.65 µm. We used a solar zenith angle of 60° to approximate a disk-average. Note that reflectances are arbitrarily scaled to provide a qualitative assessment of potential detectability.

Ozone has a number of spectral features in the IR, visible, and UV range, including a strong feature at 9.7 µm, the so-called "Chappuis" bands between 0.5-0.7 µm and the "Hartley-Huggins" bands at ~0.35-0.2 µm (Des Marais *et al.*, 2002; Domagal-Goldman *et al.*, 2014). The most promising of these from a detection standpoint is the combined Hartley-Huggins band in the near-UV centered at ~0.25 µm (Fig. 4), which is the most sensitive to small $O_2$ abundance and saturates at $O_3$ fractions of <1 ppmv. This band is also important because it lies within the proposed instrument range for space-based telescope missions currently under consideration (Dalcanton *et al.*, 2015; Mennesson *et al.*, 2016).

Peak $O_3$ abundances near the upper range of Proterozoic estimates would potentially have been a promising candidate for detection via the UV-Hartley band, but even this feature may have been challenging to observe at the lower end of $pO_2$ estimates. Combined with atmospheric $pO_2$ values orders of magnitude below this during the Archean, atmospheric $O_2$ (and possibly $O_3$) would perhaps have been difficult to detect remotely in Earth's atmosphere for most of its history. A potential exception to this early signal gap exists in the "$O_2$ overshoot" hypothesized for the Paleoproterozoic (see above), representing the fascinating possibility of a transiently detectable $O_2$-rich biogenic atmosphere during Earth's early history followed by over a billion years of predominantly undetectable $O_2/O_3$ levels. Possible shifts to more elevated $pO_2$ on shorter timescales between ~1.8 – 0.8 Ga (e.g., Planavsky *et al.*, 2014b; Sperling *et al.*, 2014; Gilleaudeau *et al.*, 2016) should also be explored for their plausibility as biosignature windows.

Methane has a number of features in the mid-IR and visible to near-IR ranges, including five weak features between 0.6-1.0 μm, two at 1.65 and 2.4 μm, and a significant feature at ~7.7-8.2 μm. The 1.65 μm $CH_4$ band would likely have been visible during Archean time (Fig. 4), but the 0.6-1.0 μm features are only expected to have appreciable depth at $pCH_4$ above ~$10^3$ ppmv (Des Marais *et al.*, 2002). In any case, in light of the model results discussed above it is possible that atmospheric $CH_4$ would have been a challenging detection for the last ~2.5 billion years of Earth's history (e.g., subsequent to the GOE), at least via low- to moderate-resolution spectroscopy. However, the very low atmospheric $pO_2$ and oceanic $SO_4^{2-}$ concentrations reconstructed for Archean time leave open the possibility that $CH_4$ would have been remotely detectable in Earth's atmosphere for the better part of Earth's first ~2 billion years (Fig. 4).

## 5. Discussion and Conclusions

By combining the geochemical reconstructions, Earth-system model results, and spectral considerations discussed above, we can place the $O_2$-$O_3$-$CH_4$ disequilibrium biosignature on Earth in a broader temporal context as a possible analog for terrestrial exoplanets. Currently available proxy and model constraints indicate that, during the Archean (~3.8 – 2.5 Ga), atmospheric $CH_4$ levels may have been generally within or above the range that would be remotely detectable. In contrast, atmospheric $O_2$ levels were many orders of magnitude below detection of either $O_2$ or $O_3$, with the possible exception of pulsed $pO_2$ increases during the late Archean (e.g., Anbar *et al.*, 2007; Kendall *et al.*, 2015). Combined proxy and model results indicate that, during the mid-Proterozoic, both $O_2$ and $CH_4$ would have been undetectable and are consistent with the possibility that $O_3$ would have been challenging to detect. An interesting exception to this may have occurred during the Paleoproterozoic, between ~2.2 – 2.0 Ga, when atmospheric $pO_2$ may have been elevated to detectable levels before decreasing again for over a billion years. Following possibly dynamic upheavals in ocean-atmosphere redox during the late Proterozoic, atmospheric $O_2/O_3$ was present at levels that would likely have been readily detectable for most of the last ~500 million years. However, results from Earth system models indicate that detection of atmospheric $CH_4$ would have been problematic with low- to moderate-resolution spectroscopy during this same period.

The realization that Earth's biosphere may have remained cryptic to conventional methods of remote detection for large periods of its history highlights the need to explore novel atmospheric biosignatures (Pilcher, 2003; Domagal-Goldman *et al.*, 2011; Seager *et al.*, 2013a) and provides further impetus for wide-ranging and systematic exploration of possible atmospheric biosignatures for application in a range of scenarios. Our results also highlight the importance of developing contextual plausibility arguments for biogenic gases in reducing atmospheres and, in particular, the presence/absence and characteristics of exoplanetary oceans. In this light, observations of time-resolved changes in atmospheric chemistry (e.g., seasonality) that can be firmly linked to quantitative models for biotic and abiotic sources will be important to consider. For example, although there may be plausible abiotic routes to high $CH_4$ in a reducing planetary atmosphere, it may be much more difficult to generate seasonal variations in atmospheric $pCH_4$ without an active surface biosphere.

Our analysis suggests that a planet with a biosphere largely (or entirely) confined to the marine realm will in many cases remain invisible to remote detection as a result of biosignature filtering by ocean biogeochemistry – a difficulty that may apply to both presence/absence and thermodynamic techniques. Our analysis suggests that the possible detection of oceans at a planet's surface (Robinson *et al.*, 2010; but see Cowan *et al.*, 2012) is a critical piece of contextual information for validating potential atmospheric biosignatures, and that planets with terrestrial biospheres (e.g., partially or entirely subaerial in scope) may be the most readily detected and characterized because of their more direct geochemical exchange with the overlying atmosphere. Ironically, in some cases planets that are very conducive to the development and maintenance of a pervasive biosphere, with large inventories of $H_2O$ and extensive oceans, may at times be the most difficult to characterize via conventional biosignature techniques.

Although our results are broadly applicable to Earth analog planets orbiting Sun-like stars, additional work will be necessary to determine the extent to which differences in the stellar environment may impact the buildup of $O_2$ and $CH_4$ (e.g., Loyd *et al.*, 2016; Meadows *et al.*, 2016), while also considering the effects of ocean chemistry, as we have done here. Nevertheless, our work further stresses the importance of including and refining UV observations for exploration of potentially habitable exoplanets, since it is quite plausible that Proterozoic Earth analogs would have detectable $O_3$ without spectrally apparent $O_2$. Moreover, high-resolution spectroscopy ($R > 20,000$), coupled with high-contrast imaging, may provide a promising avenue for detecting modern Earth-like $CH_4$ abundances on an exoplanet (e.g., Snellen *et al.*, 2015), though the feasibility of this technique has yet to be conclusively demonstrated. We would recommend investigation of this technique as perhaps the only plausible approach toward observing the $O_2$-$CH_4$ biosignature couple in true Earth analogs. More broadly, our analysis highlights the importance of including a rigorous understanding of ocean biogeochemistry into models of biosignature production and preservation on exoplanets and affirms that critical insights into the evolution of atmospheric biosignatures on Earth-like planets can be provided by better understanding Earth's dynamic history.

**Acknowledgements:** C.T.R. and T.W.L. acknowledge support from the NASA Astrobiology Institute. C.T.R. acknowledges support from the Alfred P. Sloan Foundation. E.W.S. acknowledges support from the NASA Astrobiology Institute Director's Discretionary Fund and the NASA Postdoctoral Program, administered by the Universities Space Research Association. This work benefited from the use of advanced computational, storage, and networking infrastructure provided by the Hyak supercomputer system at the University of Washington.